\documentclass[journal = nalefd,layout=twocolumn,10pt]{achemso} 
\pdfoutput=1
\usepackage{graphicx}  % needed for figures
\usepackage{bm}      % for math
\usepackage{amssymb}  % for math
\usepackage{color}
\usepackage{float}
\usepackage{subcaption}
\usepackage{braket}
\usepackage{amsmath}
\usepackage{upgreek}
\usepackage[linkcolor=blue,colorlinks=true,urlcolor=blue,citecolor=blue]{hyperref}
\usepackage[nameinlink]{cleveref}

\newcommand\amod{\alpha_\mathrm{mod}}
\newcommand\ahf{\alpha_\mathrm{hf}}

\newcommand\wmod{\Omega_\mathrm{mod}}
\newcommand\vmod{V_\mathrm{mod}}
\newcommand\wmc{\Omega_\mathrm{m}^\mathrm{cold}}
\newcommand\wmh{\Omega_\mathrm{m}^\mathrm{hot}}
\newcommand\vmc{V_\mathrm{m}^\mathrm{cold}}
\newcommand\vmh{V_\mathrm{m}^\mathrm{hot}}
\newcommand\vout{V_\mathrm{out}}
\newcommand\vhf{V_\mathrm{hf}}
\newcommand\whf{\Omega_\mathrm{hf}}
\newcommand\ts{\uptau_\mathrm{s}}

\newcommand\pin{P_\mathrm{in}}

\usepackage{textgreek}

\usepackage{ulem}

\crefname{section}{Sec.}{Secs.}

\title{Vibrational resonance amplification in a thermo-optic optomechanical nanocavity}

\author{Guilhem Madiot}
\affiliation{Centre de Nanosciences et de Nanotechnologies, CNRS,  Université Paris-Saclay, Palaiseau, France}

\author{Sylvain Barbay}
\affiliation{Centre de Nanosciences et de Nanotechnologies, CNRS,  Université Paris-Saclay, Palaiseau, France}

\author{Rémy Braive}
\affiliation{Centre de Nanosciences et de Nanotechnologies, CNRS,  Université Paris-Saclay, Palaiseau, France}
\alsoaffiliation{Université de Paris, F-75006 Paris, France}
\email{remy.braive@c2n.upsaclay.fr}

\begin{document}

\begin{abstract}
\textbf{Vibrational resonance amplifies a weak low-frequency signal by use of an additional non-resonant high-frequency modulation. The realization of weak signal enhancement in integrated nonlinear optical nanocavities is of great interest for nanophotonic applications where optical signals may be of low power.
Here, we report experimental observation of vibrational resonance in a thermo-optically bistable photonic crystal optomechanical resonator with an amplification up to +16 dB. The characterization of the bistability can interestingly be done using a mechanical resonance of the membrane, which is submitted to a strong thermo-elastic coupling with the cavity. }
\end{abstract}
  
Phase transitions and double-well potentials have been extensively exploited to amplify or detect weak signals \cite{rajasekar2016nonlinear}. Such general physical concepts, observed in various fields of science are at the heart of the phenomenon of vibrational resonance \cite{Landa_2000} (VR). Introduced as an analogy with the well known stochastic resonance \cite{Dykman1995StochasticRI}, VR uses a high-frequency (HF) periodic signal to amplify a low-frequency (LF) input signal.
It has been theoretically studied in different types of nonlinear systems, e.g. in neural network \cite{Deng2010}, in excitable systems \cite{ZaikinPRE} or in biological networks \cite{RAJASEKAR20123435}. Several experimental demonstrations have also been conducted in electronic circuits \cite{ULLNER2003348}, bistable VCSELs \cite{PRLVR,PhysRevE.89.062914} or electromechanical Duffing resonator \cite{chowdhury2019weak}, for example. Despite the growing interest for photonic nanodevices and their use for signal processing and sensing, VR has not yet been demonstrated using nonlinear optical nanocavities.

Among optical nonlinearities, the thermo-optic effect overwhelmingly manifests itself in photonic micro and nano-devices. Resulting from the temperature dependence of a material refractive index, it is commonly utilized to introduce tunability and to realize elementary computational all-optical components, such as photonic switches \cite{watts2013adiabatic}, phase-shifter \cite{harris2014efficient} or integrated spectrometers \cite{souza2018fourier}.
Due to the ultimate electromagnetic field confinement achievable in optical nanocavities, strong thermo-optic effects can lead to multistability \cite{Uesugi:06,Gao:17,PhysRevApplied.14.024073}, which has been widely used to study nonlinear dynamical behaviors such as, e.g., self-stability \cite{Carmon:04}, self-pulsing \cite{Baker:12}, excitability \cite{PhysRevLett.97.143904} or soliton modelocking \cite{Joshi:16}. The usefulness of optical nanocavities has also largely been demonstrated to sense or manipulate mechanical vibrations \cite{aspelmeyer2014cavity}. Thus complex dynamics can emerge due to a combination of thermo-optic and optomechanical nonlinearities enabling e.g. utilization in sensing applications \cite{Deng:13},  chaos generation \cite{navarro2017nonlinear}, or electro-optomechanical self-oscillation \cite{Allain2021} .
%\cite{Krause2015}

\begin{figure*}[!ht]
\begin{center}
\includegraphics[scale=0.55]{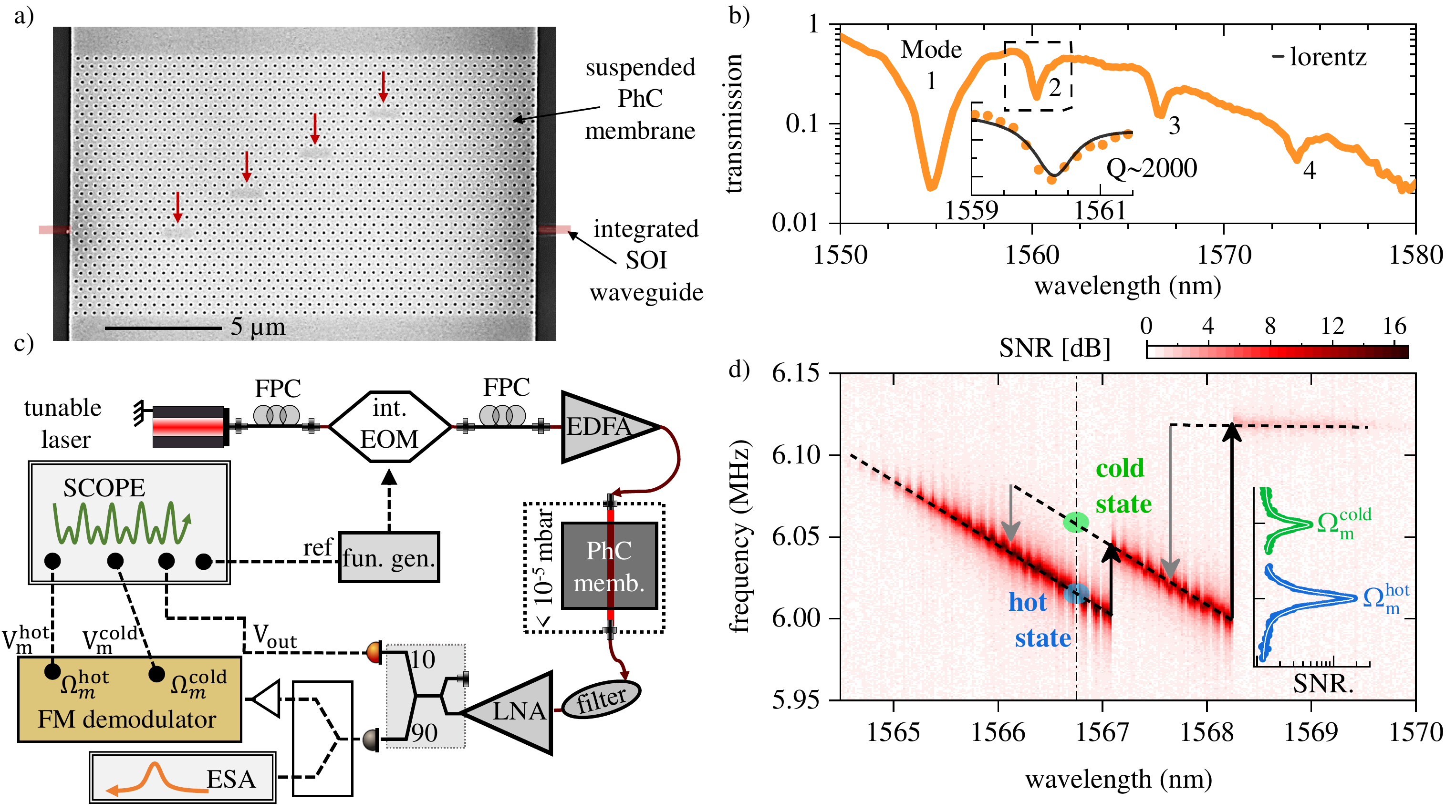}
\end{center}
{\phantomsubcaption\label{fig1:a}}
{\phantomsubcaption\label{fig1:b}}
{\phantomsubcaption\label{fig1:c}}
{\phantomsubcaption\label{fig1:d}}
\caption{a) SEM image of the suspended photonic molecule -- made of L3-defect cavities (arrows) -- and underneath SOI waveguide (red stripe). b) Waveguide transmission spectrum with four dips corresponding to the PhC molecule photonic eigenmodes. c) Experimental setup. d) Noise spectrum of the output optical field as a function of the injected laser wavelength. Scan performed over the 2\textsuperscript{nd} and 3\textsuperscript{rd} photonic modes. The fundamental mechanical resonance experiences a photothermal shift proportional to the intracavity energy. The black (resp. gray) arrows indicate the abrupt jumps of the mechanical frequency observed at the bistabilities' edges under upward (backward) scan of the laser wavelength. inset: noise spectra taken when the optical resonator is the hot state (blue) and in the cold state (blue). The mechanical quality factor $Q_m\sim1400$ is estimated from the Lorentzian fits (white).}
\end{figure*}

While stochastic resonance has been observed in optomechanical systems \cite{PhysRevA.79.031804,monifi2016optomechanically}, VR remains unexplored both theoretically and experimentally. Within this framework, VR in optical nanocavities supporting simultaneously mechanical modes and nonlinearities is a key development for potential nanophotonic and optomechanical applications.

In this letter, we demonstrate VR amplification of a weak LF signal in a suspended thermo-optic photonic crystal (PhC) nanoresonator in the bistable regime. Thanks to a strong thermo-elastic coupling between one mechanical resonance of the PhC membrane and the optical cavity load, the state of the thermo-optic cavity can be readout through the mechanical mode spectral position. It further enables to characterize the threshold modulation amplitude below which the modulation signal does not trigger switching between the states of the bistability. Then by adding a non-resonant HF signal, we observe up to $+16$ dB amplification of the weak LF signal. %This work constitutes an unprecedented observation of VR in an optical microcavity. 

The system is a $12\times21$ $\upmu$m$^2$ rectangular InP membrane with thickness $265$ nm suspended over a $250$ nm airgap. The membrane is pierced with a 2D PhC in which four L3 defect-cavities are diagonally arranged (\cref{fig1:a}) such that the electromagnetic field confined in each defect can evanescently couple to the neighboring ones \cite{Chalcraft:11}. A SOI waveguide is placed underneath the membrane for integrated light injection \cite{tsvirkun2015integrated}.  The electromagnetic modes confined in the PhC molecule are characterized by injecting a broadband superluminescent diode through the waveguide. The IR spectral analysis of the waveguide transmission, shown in \cref{fig1:b}, evidences four resonance dips.
Here, the use of a PhC molecule enables higher optomechanical coupling to the MHz drum modes of the membrane -- i.e. whose associated displacement field is extended on the whole membrane -- and on which we focus here. 
 
The PhC eigenmodes can alternatively be injected with a preliminary amplified tunable laser (see \cref{fig1:c}) enabling coherent and resonant excitation of the photonic modes. Doing so, the brownian motion of the suspended membrane is optomechanically transduced into the optical field. 
 The analysis of the output field noise spectrum is done via an electrical spectrum analyzer (ESA). The membrane carries several mechanical modes in the MHz domain. Focusing on the $6$ MHz fundamental mechanical mode, we plot the noise spectrum of the output optical field as a function of the input laser wavelength in \cref{fig1:d}. The laser is scanned over the 2\textsuperscript{nd} and 3\textsuperscript{rd} photonic modes and the estimated power sent to the photonic molecule is $\pin=2.4$ mW. The use of sufficiently high circulating power reveals a thermo-optic nonlinearity which leads both photonic modes to shift and to become bistable. Note that the 2\textsuperscript{nd} mode experiences a stronger red-shift than the 3\textsuperscript{rd} mode. 
 The thermal origin of this nonlinear effect is confirmed in the following by the temporal response of the optical cavity. Interestingly, the intra-cavity field intensity is partially absorbed which also provokes the photothermal shift of the mechanical frequencies through the thermo-elastic effect.
Indeed, as the laser wavelength is upwardly swept from off-resonance to on-resonance the mechanical frequency experiences a linear photothermal shift. The abrupt intra-cavity intensity drop occurring when passing the edge of the first bistability (2\textsuperscript{nd} photonic resonance) translates into an abrupt red-shift of the mechanical frequency (black arrow). The same phenomenon happens when scanning the 3\textsuperscript{rd} photonic bistable resonance (\cref{fig1:d}). 
The downward scan can only be performed manually and gives access to the hysteresis turning point. The backward jumps of the mechanical frequency are measured and shown with the gray arrows. The corresponding trajectory superposes with the upward trajectory except in the bistable regions. Therefore it is possible to evaluate the intra-cavity field intensity by measuring the frequency shift of a mechanical mode.  In the following we set the laser wavelength at the center of the second photonic mode bistability ($\lambda = 1566.75$ nm) to symmetrize the double-well potential \cite{chowdhuryPRL}. Thus we name hot (blue) and cold (green) mechanical states the corresponding mechanical frequencies $\wmh/2\pi$ and $\wmc/2\pi$ (see inset in \cref{fig1:d}). A lorentzian fit of the mechanical lineshape (white lines) returns a mechanical linewidth $\Gamma_m/2\pi=4.3$ kHz, hence a mechanical quality factor of about 1400.

\begin{figure}[!h]
\begin{center}
\includegraphics[scale=1]{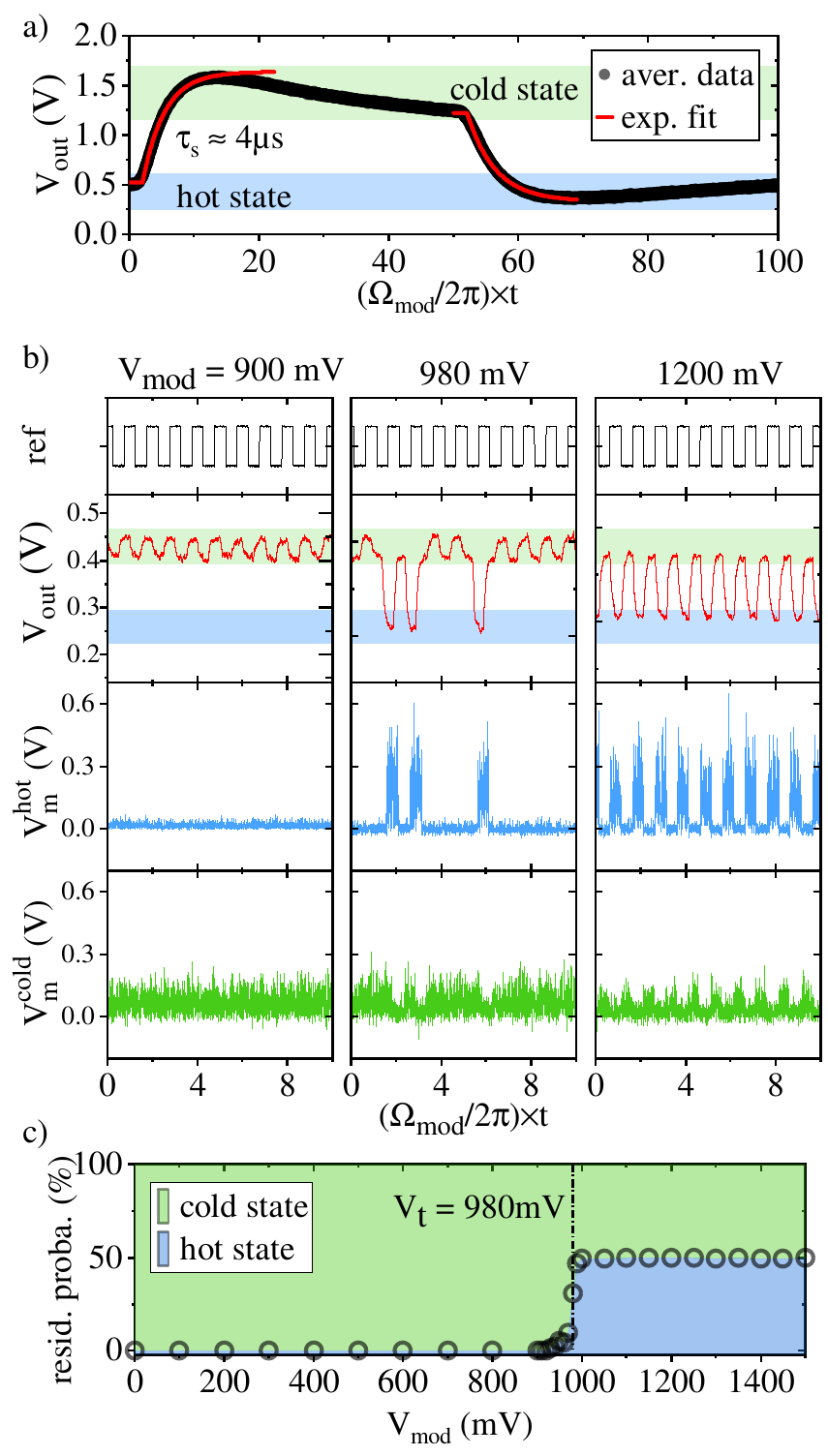}
\end{center}
{\phantomsubcaption\label{fig2:a}}
{\phantomsubcaption\label{fig2:b}}
{\phantomsubcaption\label{fig2:c}}
\caption{a) Output signal ($\vout$) averaged during hundreds of modulation periods with $\wmod/2\pi=10$ kHz. The switching decay times is fitted with exponential fits. b) For $\vmod=900$, $980$ and $1200$ mV, from top to bottom: reference modulation signal, transmitted signal $\vout$ and demodulated noise amplitude at the hot (cold) state mechanical frequency $\wmh$ ($\wmc$). $\wmod/2\pi=10$ Hz. c) Probability for the optical state to be in the hot (blue) or cold (green) state.}
\end{figure}

When the optical resonator state flips, it requires a certain time to reach its new stable state.
Under strong modulation of the input field intensity, the optical state is asked to flip twice per modulation period, which is possible only if the switching time $\ts$ is short enough. Therefore, the maximum modulation frequency allowing the system to accurately flip is given by $2\ts^{-1}$. 
We estimate this cut-off modulation frequency by measuring the typical switching time of the system. To do so, we modulate the input laser field slowly enough such that the transitory regime -- in which the system leaves a stable state to reach the other one -- can be measured. The laser wavelength is modulated in an intensity electro-optic modulator (EOM, with half-wave voltage $V_\pi=7.5 V$) on which we apply a square electrical signal carrying amplitude $\vmod$ and frequency $\wmod$. 
A femtowatt sensitivity photodetector converts the transmitted laser field into a voltage $\vout$ we input in an oscilloscope (see \cref{fig1:c}).

Modulating the input field with $\vmod=1$ V and $\wmod/2\pi=10$ kHz, the system response is recorded over several hundred of modulation cycles. The data are averaged cycle by cycle and plotted in \cref{fig2:a}. During one cycle, the resonator, initially set in the hot state, transits towards the cold state and then switches back to the hot state at half a modulation cycle. Note that since the cavity load is accessed through the waveguide transmission, the cavity hot state (resp. cold state) corresponds to the low (high) transmission level.
This transient regime manifests as an exponential decay from one state to the other. Fitting the averaged data provides the switching time $\ts\approx4$ $\upmu$s. This value is similar for both transitions hot$\rightarrow$cold or cold$\rightarrow$hot. The corresponding cut-off frequency, as discussed above, is of the order of 125 kHz, which is consistent with previous estimation made in a similar InP photonic crystal nanocavity \cite{Brunstein}.

In order to clearly resolve the mechanical frequency, we use in the following a LF weak-signal with frequency of $\wmod/2\pi=10$ Hz. In principle, the use of faster signal should not prevent from the realization of amplification as demonstrated below. However, the modulation tones present in the driving field couple with the optomechanical cavity such that they are transduced into the mechanical frequency domain \cite{pelka2021floquet}. As a result, the mechanical response at resonance would be weakened as sidebands would be created. % the mechanical linewidth.
Keeping $\wmod\ll\Gamma_m$ guarantees here the demonstration of VR.

Once the laser wavelength and modulation frequency are set, the switching between optical stable states can be achieved only if the modulation amplitude passes a certain threshold. The demonstration of VR amplification relies on the use of a weak signal, i.e. with amplitude lying below this threshold, whose knowledge is consequently required.
To evaluate the latter, we increase the modulation amplitude and check the optical response. When the amplitude is sufficiently high, the optical state flips periodically at the modulation frequency.

In our case, the optical state can be accessed both through the transmission signal or through the spectral position of the mechanical resonance. In parallel of the direct output measurement ($\vout$), the output signal is analyzed in the ESA via a fast photodetector. From the spectrum, we determine the two positions of the mechanical peak $\wmc/2\pi=6.0572$ MHz and $\wmh/2\pi=6.0164$ MHz for the cold and hot optical states, respectively.
The RF signal is therefore amplified and demodulated using a lock-in amplifier. Two demodulation channels are used at frequencies $\wmc$ and $\wmh$ and with 1 kHz wide passband filters. The two corresponding demodulated signal amplitudes $\vmc$ and $\vmh$ are recorded in the oscilloscope. Due to the demodulation filter whose bandwidth is smaller than the mechanical linewidth, noise in the demodulated signal is induced by frequency fluctuations of the mechanical resonance.

The resulting time traces are shown for three values of the modulation amplitudes in \cref{fig2:b}. For each situation, we show the reference modulation signal (black), the transmission signal (red) and the demodulation amplitudes (blue and green). 
Around a threshold value $V_t=980$ mV, the optical state starts to flip but tends to remain in the cold state. The dynamics of the mechanical frequency is in perfect correlation with these jumps. For $\vmod>V_t$, the optical state jumps are perfectly synchronous with the modulation reference signal.
Note that the signal-to-noise ratio in the mechanical spectrum is higher in the hot state, due to higher optomechanical coupling which is proportional to the energy stored in the optical cavity. This difference is also visible in the amplitude of the mechanical response (inset of \cref{fig1:d}). 

For a given time trace, one can calculate the residence probability of the optical state. It consists in evaluating the amount of time spent by the system in the cold (or hot) optical state. To do so, a threshold line is arbitrarily chosen in between the two corresponding amplitude levels. The probability for the system to set in the initial state, is 100\% for low modulation amplitudes (see \cref{fig2:c}). It quickly converges to 50\% around  the threshold amplitude $V_t = 980$ mV. Such data treatment performed on $\vout$, $\vmh$ or $\vmc$ all provide the same residence probability shown in \cref{fig2:c}.

\begin{figure}[!h]
\begin{center}
\includegraphics[scale=1]{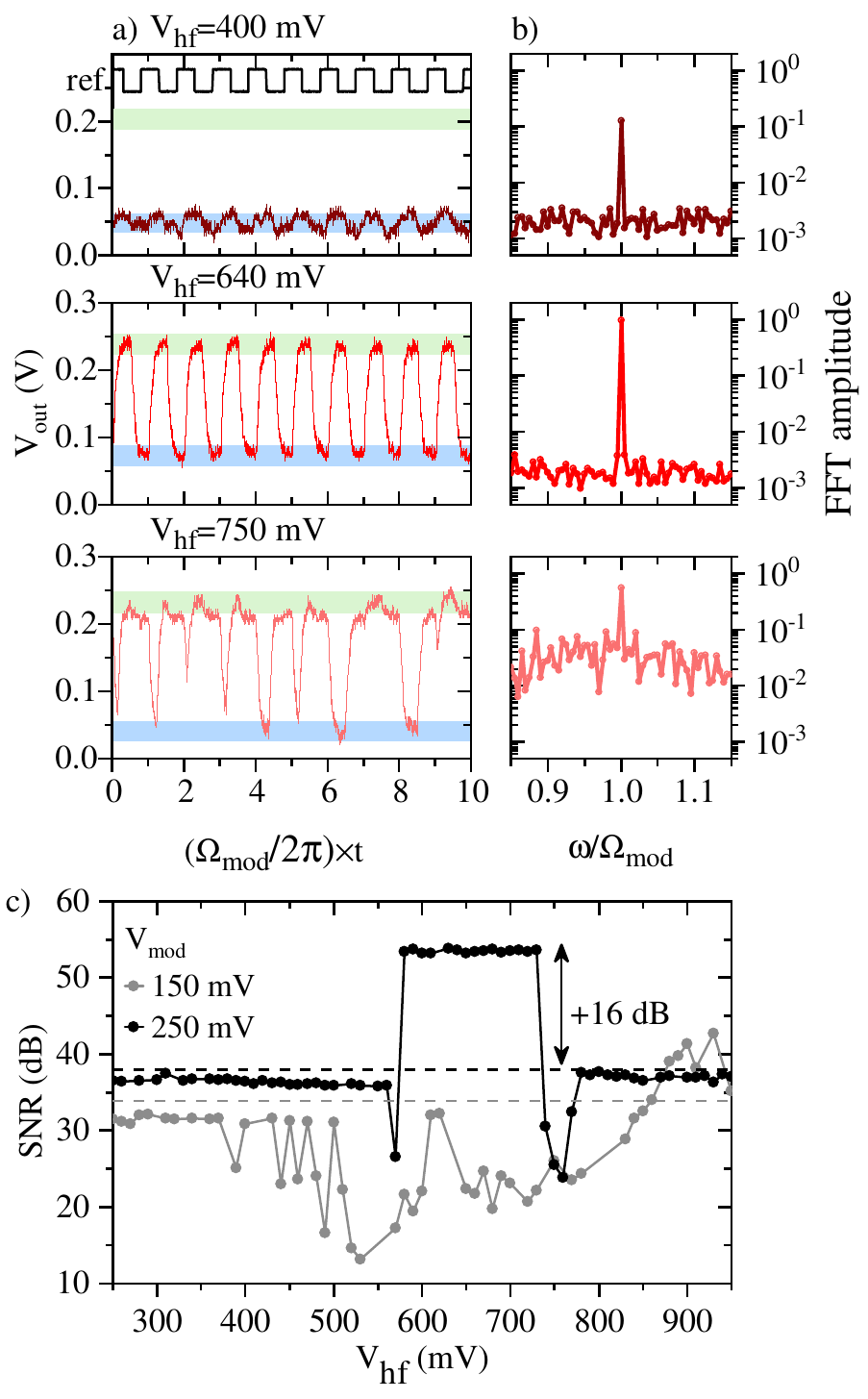}
\end{center}
{\phantomsubcaption\label{fig3:a}}
{\phantomsubcaption\label{fig3:b}}
{\phantomsubcaption\label{fig4}}
\caption{a) Time-domain and b) frequency-domain transmitted signal $\vout$ for three values of the high frequency signal amplitude $\vhf=400$, $640$ and $750$ mV. c) SNR as the function of the HF signal below and above the amplification threshold, resp. with $\vmod=150$ mV (gray) and $\vmod=150$ mV (black). We use $\whf/2\pi=80$ kHz.}
\end{figure}

With the full knowledge on the cut-off frequency and the threshold modulation amplitude, we investigate the amplification of a weak signal via an additional HF intensity modulation with frequency $\whf$ and amplitude $\vhf$. Thus the field now carries both tones $\wmod$ and $\whf$:
\begin{equation*}
V_\mathrm{ref} = \vmod \text{sign}(\sin(\wmod t) + \vhf\cos(\whf t)
\end{equation*} 

When $\vmod<V_t$ and $\vhf=0$, we know that the square signal cannot flip the optical state. However, the increase of $\vhf$ comes with a significant distortion of the bistability until the switching process can be re-activated. Ensuring that $\wmod\ll\whf$ \cite{chowdhury2019weak}, we arbitrarily set the frequency of the HF signal to $\whf/2\pi=80$ kHz and $\vmod=250$ mV, and record the output optical response for increasing value of $\vhf$. The results are shown through three examples. In each case, we plot the time trace (\cref{fig3:a}) and the corresponding FFT spectrum (\cref{fig3:b}). At frequency $\wmod$, a peak testifies the presence of the LF square modulation. Its amplitude increases when the switching process becomes more efficient.
While at $\vhf=400$ mV the system remains in the hot state, switching is obtained at $640$ mV, with 100\% fidelity to the LF signal. In both cases, the noise remains constant in the FFT spectrum. At $\vhf=750$ mV, the fidelity drops which results from a clear increase of the noise floor.

For a given LF signal amplitude $\vmod$, the signal-to-noise ratio (SNR) is evaluated as a function of the HF signal amplitude $\vhf$ \cite{PRLVR}. The SNR is taken as the amplitude of the LF modulation peak in the FFT spectrum, to the noise level around this peak. This measurement is reproduced for several values of $\vmod$. The resulting amplification curves are shown in \cref{fig4}. Amplification occurs when the SNR is higher than its value for $\vhf=0$. These reference levels are indicated with dashed lines.
% (: 38.96 dB; gray: 33.97 dB). 
We observe a significant amplification ($\vmod=250$ mV, black curve) up to $+16$ dB. The range of $\vhf$ -- in which amplification is observed -- is delimited by transitory regions where the switching is imperfectly achieved, resulting in a high noise level, as discussed above.
Weakening further the LF signal amplitude modifies amplification features as observed with the gray curve in \cref{fig4}, where we use $\vmod=150$ mV. An amplification of 5 dB is observed around $\vhf=900$ mV. It is likely that higher amplification could be obtained by increasing $\vhf$.

The observation of VR reported here can be consistently modeled with a single waveguide-coupled optical mode including a photothermal nonlinearity. Numerical simulations, available in the Supporting information, provide a qualitative agreement with the experimental findings.

VR amplification could not be performed by exploiting the demodulated signals, $\vmh$ and $\vmc$ at the mechanical resonances because of the appearance of sidebands through the HF tone. Indeed, the presence of the latter (with $\Omega_m>\whf\gg\Gamma_m$) can produce strong sidebands reducing the effective mechanical response at resonance, and whose intensity depend on the modulation depth and frequency \cite{pelka2021floquet}. Therefore, optimal adjustment of these parameters is expected to render possible the characterization of VR in the mechanical domain.
Other solutions might also be investigated. Self-sustained oscillation regimes enabled in high optical quality factor and low phase-noise optomechanical cavities \cite{Ghorbel} would constitute an ideal support to explore manipulation of mechanical oscillations through the presently described phenomenon of vibrational resonance.

In conclusion, we have demonstrated amplification of a weak LF signal using vibrational resonance in a waveguide-coupled thermo-optic nano-resonator. The characterization of the bistable system is made both through the waveguide transmission optical field and through the thermo-mechanical effect induced by photo-thermal absorption in the optomechanical cavity. 
This strong thermo-elastic coupling could be exploited in sensing applications \cite{Deng:13} or in the realization of FM modulation of nanomechanical oscillators, as demonstrated here. Our system made of coupled PhC cavity constitutes an ideal test-bed for such investigations.  Additionally, the amplification of weak signals could be enhanced and enriched by the presence of multistability, which can occur when two or more bistable resonances overlap \cite{PhysRevApplied.14.024073}.

\paragraph{Acknowledgement}
%\textbf{acknowledgments:}
This work is supported by the French RENATECH network, the European Union’s Horizon 2020 research innovation program under grant agreement No 732894 (FET Proactive HOT), the Agence Nationale de la Recherche as part of the “Investissements d’Avenir” program (Labex NanoSaclay, ANR-10-LABX-0035) with the flagship project CONDOR and the JCJC project ADOR (ANR-19-CE24-0011-01).

\appendix
\section{Supporting information}
\label{SI}
\subsection{Theoretical model}

Here we derive a model to describe the vibrational resonance amplification of a weak signal in a thermo-optic waveguide-coupled optical mode. We consider a single optical mode with complex amplitude $a$. We write $\omega_c$, $\kappa_i$, $\kappa_e$, $s_i$ and $\omega_L$ respectively the mode frequency, its intrinsic loss rate, its external loss rate, the laser amplitude and the laser frequency.
The system can be modelled within coupled mode theory approach:
\begin{equation}
\label{eq1}
    \dot{a}(t) = \Big(j(\omega_L-\omega_c) - \frac{\kappa_t}{2}\Big)a(t) + \sqrt{\kappa_e}\times s_i(t)
\end{equation}
where $\kappa_t=\kappa_i+\kappa_e$ is the optical mode total resonance linewidth.

We introduce the thermo-optic nonlinearity through the cavity temperature dynamics, with $\Delta T$ the temperature elevation in the optical cavity:
\begin{equation}
\label{eq2}
    \dot{\Delta T} = \mathcal{K}_t\Big(\kappa_{th}|a|^2 - \Delta T\Big)
\end{equation}
with $\mathcal{K}_t$ the thermalization rate of the cavity and $\kappa_{th}$ the thermal resistance (in K/J).

The optical mode resonance wavelength is thermally shifted by an amount:
\begin{equation}
\label{eq3}
\Delta\lambda = \frac{\lambda_0}{n_0}\frac{d\text{n}}{dT} \Delta T
\end{equation}

with $\lambda_0 = 2\pi c/\omega_0$ the optical mode wavelength -- $\omega_0$ and $n_0$ being respectively the cavity natural frequency and refractive index, both taken at room temperature -- and $\frac{d\text{n}}{dT}$ the thermo-optic coefficient of the material.

\begin{table}
 \begin{center}
% \vspace{-2cm}
 \begin{tabular}{ p{5cm} p{2.8cm}}
 \hline 
 Physical parameter & value\\
 \hline
 $\lambda_0$            : resonance wavelength      & 1566.3 nm\\
 $\kappa_i$             : internal loss rate   & 28.2 GHz\\
 $\kappa_e$             : external loss rate   & 18.4 GHz\\
 $n_0$                  : refractive index              & 3.16 \\
 $\frac{d\text{n}}{dT}$ : thermo-optic coef. & $1.9298\cdot10^{-4} \text{K}^{-1}$\\
 $\mathcal{K}_t$        : Thermalization rate               & 125 kHz\\
 $\kappa_{th}$          : thermal resistance                & 1.62 $\text{K.fJ}^{-1}$
\end{tabular}
\end{center}
\caption{List of parameters used in the numerics.}
\label{AllVars}
\end{table}

In the numerical simulation, we integrate \cref{eq1,eq2,eq3} using the real and imaginary parts of $a$, and considering a temperature dependent resonance frequency for the optical mode:

\begin{equation}
\omega_c = \omega_0 (1 - \frac{1}{n_0}\frac{d\text{n}}{dT}\Delta T)
\end{equation}

The ODEs are integrated after time-normalization $t\longrightarrow \mathcal{K}_t t$, as $\mathcal{K}_t$ constitutes the cut-off frequency of the thermo-optic nonlinearity. The input field is modelled with a square modulation at $\wmod$ plus a high frequency sinusoidal signal at $\whf$. Both components have respective amplitudes given by the modulation depths, respectively $\amod$ and $\ahf$:

\begin{align}
s_i(t) = \sqrt{\pin}\Big[ 1 + e^{j\pi\big(-\frac{1}{2} + \amod f(\wmod t) + \ahf\cos(\whf t) \big)} \Big] 
\end{align}

where $\pin$ is the laser input power and $f(t)$ is the square signal function,
\begin{align*}
f(t)=\frac{4}{\pi}  \sum_{p=0}^N \frac{\sin\big((2p+1)t\big)}{2p+1}
\end{align*}

In the numerics we use N=15. All the constants are given in \cref{AllVars}. 

\subsection{Numerical simulations}

\begin{figure*}[htp]
\begin{center}
\includegraphics[scale=1]{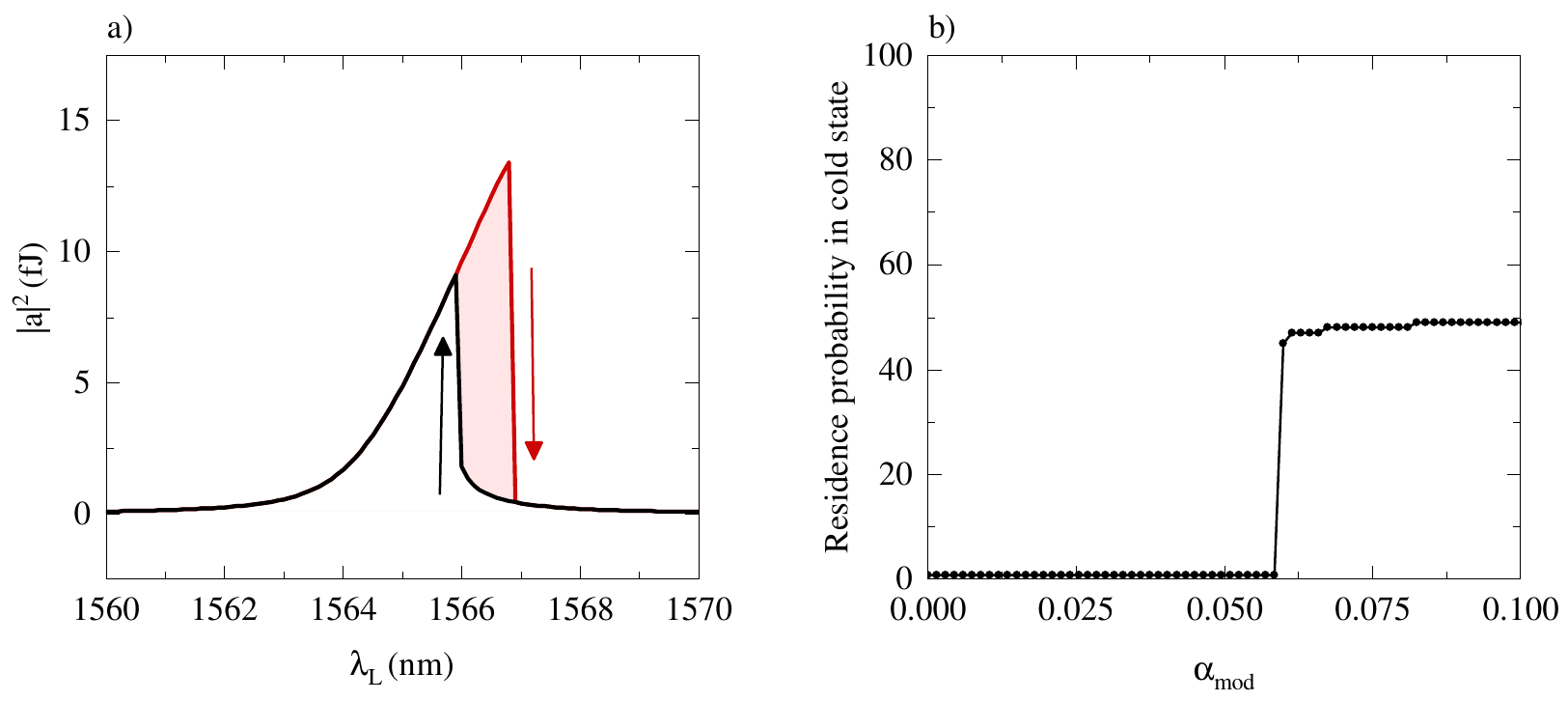}
\end{center}
{\phantomsubcaption\label{simuFig1:a}}
{\phantomsubcaption\label{simuFig1:b}}
\caption{a) We evaluate the mode energy $|a|^2$ while the laser wavelength $\lambda_L = 2\pi c/\omega_L$ is swept forward (red) and backward (black), using $\pin=10$ mW, giving rise to a $\sim1$ nm wide bistability (red area). $\amod=\ahf=0$. b) Residence probability as a function of the modulation depth for $\lambda_L=1566.30$ nm and $\wmod/2\pi=500$ Hz.}
\end{figure*}

In \cref{simuFig1:a} the mode intensity is numerically evaluated as a function of the laser wavelength under forward and backward scans (resp. black and red curves).
Tuning the input power $\pin$, the spectral span ($\sim1$ nm) of the bistability (red region) is adjusted to qualitatively match the experiment. Note that playing with the power of with the thermo-optic coefficient is equivalent in this model. We find the best agreement for $\pin=10$ mW.
%For three position of the laser in the bistability, we evaluate the residence probability for the cavity to be in the cold state as a function of the modulation depth $\alpha_\text{mod}$ (see \cref{simuFig1:b}, right). Here we use $\wmod/2\pi=500$ Hz. 

Using the laser wavelength $\lambda_L=1566.3$ nm), we numerically resolve the ODE for increasing modulation depth $\amod$. The modulation frequency is set to $\wmod/2\pi=500$ Hz rather than 10 Hz in the experiments, in order to reduce the integration time. The time trace $|a|^2(t)$ is used to calculate the residence probability of the optical mode (see \cref{simuFig1:b}).

\begin{figure}[htp]
\begin{center}
\includegraphics[scale=1]{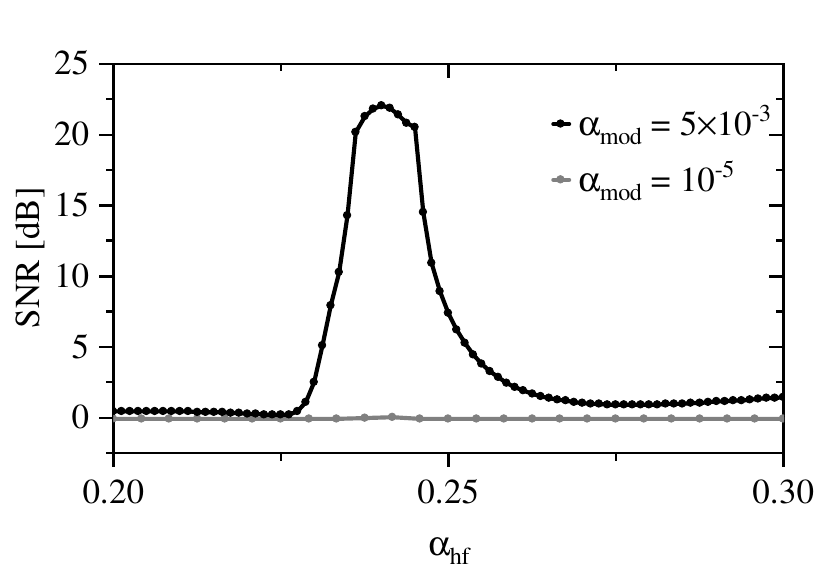}
\end{center}
\caption{Signal-to-noise ratio (SNR) plotted as a function of the high-frequency signal amplitude, $\ahf$, for $\amod=0.005$ (black) and  $\amod=10^{-5}$ (gray).  $\wmod/2\pi=500$ Hz and $\whf/2\pi=10$ kHz.}  
\label{fig2}
\end{figure}

Enabling the high-frequency modulation with $\whf/2\pi=10$ kHz, the amplification curve is obtained by evaluating the signal-to-noise ratio (SNR) of the weak low-frequency modulation in the mode response, as a function of the HF signal amplitude, $\ahf$. The results are shown for two different values of $\amod$, in \cref{fig2}. Significative amplification is obtained for $\amod=0.005$ (black curve), but not for $\amod=10^{-5}$ (gray curve). The shape of the amplification curve as well as its amplitude of nearly 22 dB show a good qualitative agreement with the experimental results. %Note that in the frame of this model we indeed expect higher amplification since the amplitude difference between the two stable states is over-estimated when neglecting the presence of 

Note that the model here accounts for a single optical mode. Therefore the significative overlap between bistable optical modes that we observe experimentally is not captured by the present simple model.

%\begin{footnotesize}
\bibliography{ThermoOptic_VR_bib}
%\end{footnotesize}

\end{document}